\documentstyle[prl,aps,psfig]{revtex}
\begin {document}

\draft

\twocolumn
[\hsize\textwidth\columnwidth\hsize\csname @twocolumnfalse\endcsname

\title{The Magnetic Spin Ladder  
(C$_{5}$H$_{12}$N)$_{2}$CuBr$_{4}$:
High Field Magnetization  and  Scaling Near Quantum Criticality} 
\author{B. C. Watson, V. N. Kotov, and M. W. Meisel}
\address{Department of Physics and The Center for Condensed Matter  
Sciences, \linebreak 
University of Florida, 
\mbox{P.O. Box 118440}, 
Gainesville, FL 32611-8440.}

\author{D. W. Hall}
\address{National High Magnetic Field Laboratory, 
Florida State University, Tallahassee, FL 32310.}

\author{G. E. Granroth, W. T. Montfrooij, and S. E. Nagler}
\address{Oak Ridge National Laboratory, Building 7692, 
MS 6393, \mbox{P.O. Box 2008}, Oak Ridge, TN 37831.}

\author{D. A. Jensen, R. Backov, M. A. Petruska, G. E. Fanucci, and D. R. Talham}
\address{Department of Chemistry, University of Florida, 
\mbox{P.O. Box 117200}, 
Gainesville, FL 32611-7200.}

\date{\today}

\maketitle

\begin {abstract}
The magnetization, $M(H \leq 30$ T, 0.7 K $\leq T \leq 300$ K), of 
(C$_{5}$H$_{12}$N)$_{2}$CuBr$_{4}$ has been 
used to identify this system as an $S=1/2$ Heisenberg 
two-leg ladder in the strong-coupling limit, 
$J_{\perp} =  13.3$ K and $J_{\parallel} = 3.8$ K, with 
$H_{c1} = 6.6$ T and $H_{c2} = 14.6$ T.  
An inflection point in $M(H, T = 0.7$ K) at half-saturation, $M_{s}/2$, 
is described by an effective \emph{XXZ} chain.  The data 
exhibit universal scaling behavior in the vicinity of $H_{c1}$ and $H_{c2}$, 
indicating the system is near a quantum critical point.
\end {abstract}


\twocolumn
]

Magnetic spin ladders are a class of low dimensional materials with structural 
and physical properties between those of 1D chains and 2D planes.  
In a spin ladder, the vertices possess unpaired spins 
that interact along the legs via $J_{\parallel}$ and along the rungs 
via $J_{\perp}$, but are isolated from 
equivalent sites on adjacent ladders, \emph {i.e.} 
interladder $J^{\prime} \ll J_{\parallel} , J_{\perp}$.  
Recently, a considerable amount of attention has 
been given to the theoretical and experimental investigation of spin 
ladder systems as a result of the observation that the 
microscopic mechanisms in these systems may be related to the 
ones governing high temperature superconductivity \cite{Dagotto,Sachdev1}.
The phase diagram of the antiferromagnetic spin ladder
in the presence of a magnetic field is particularly interesting.
At $T=0$ with no external applied field, the ground state is a gapped, 
disordered quantum
spin liquid. At a field $H_{c1}$, there is a transition to a gapless 
Luttinger liquid phase,
with a further transition at $H_{c2}$ to a fully polarized state.  
Both $H_{c1}$ and $H_{c2}$
are quantum critical points \cite{Sachdev1}.  Near these points, the 
magnetization has been
predicted to obey a universal scaling function \cite{Sachdev2}, but until now, this 
behavior has not been observed experimentally.
	
A number of solid-state materials have been proposed as examples of spin 
ladder systems, and an extensive set of experiments have been 
performed on Cu$_2$(C$_5$H$_{12}$N$_{2}$)$_{2}$Cl$_{4}$, 
referred to as Cu(Hp)Cl \cite{Chiari}. The initial work identified this material as 
a two-leg $S = 1/2$ spin ladder 
\cite{Chiari,Hammar1,Hayward,Chaboussant1,GC,Weihong,Chaboussant2,Chaboussant3,Chaboussant4}.  
Although quantum critical behavior has been preliminarily identified in this system 
near $H_{c1}$, this assertion is based on the use of scaling parameters identified 
from the experimental data rather than the ones predicted 
theoretically \cite{Chaboussant3,Chaboussant4}.  
Furthermore, more recent work has debated the 
appropriate classification of the low temperature properties 
\cite{Hammar2,Elstner,Calemczuk,Ohta,Wang,Mayaffre,Stone}.  
Clearly, additional physical 
systems are necessary to experimentally test the predictions of the 
various theoretical treatments of two-leg $S=1/2$ spin ladders.

Herein, we report evidence that identifies bis(piperi-
dinium)tetrabromocuprate(II), 
(C$_{5}$H$_{12}$N)$_{2}$CuBr$_{4}$ \cite{Patyal}, 
hereafter referred to as BPCB, 
as a two-leg $S=1/2$ ladder that 
exists in the strong-coupling limit, $J_{\perp}/J_{\parallel} > 1$.  
High-field, low-temperature magnetization, $M(H \leq 30$ T, $T \geq 0.7$ K), 
data of single crystals and powder samples have been fit to obtain  
$J_{\perp} =  13.3$ K, $J_{\parallel} = 3.8$ K, and $\Delta \sim 9.5$ K, 
\emph{i.e.} at the lowest temperatures 
finite magnetization appears at $H_{c1} = 6.6$ T 
and saturation is achieved at $H_{c2} = 14.6$ T. 
An unambiguous inflection point in the magnetization, $M(H,T=0.7$ K), 
and its derivative, $dM/dH$, is observed at half the 
saturation magnetization, $M_{s}/2$. 
This feature is symmetric about $M_{s}/2$, consistent with expectations for 
a simple spin ladder.  Any presence of asymmetry, as was observed in 
Cu(Hp)Cl \cite{Hammar1,Hayward,Chaboussant1,GC}, most likely arises from 
other factors.  Our $M_{s}$/2 feature cannot be explained 
by the presence of additional  
exchange interactions, \emph{e.g.} diagonal frustration $J_{F}$, but  
is well described by an effective \emph{XXZ} chain, onto which the 
original spin ladder model (for strong-coupling) 
can be mapped in the gapless regime $H_{c1} < H < H_{c2}$ \cite{Totsuka}.  
After determining $H_{c1}$ and with no additional adjustable parameters, 
the  magnetization data are observed to
obey a universal scaling  function \cite{Sachdev2}.  
This observation supports 
our identification of BPCB as a two-leg $S=1/2$ Heisenberg spin ladder with 
$J^{\prime} \ll J_{\parallel}$.  

The crystal structure of BPCB has been determined 
to be monoclinic with stacked pairs of $S = 1/2$ Cu$^{2+}$ ions forming magnetic dimer 
units \cite{Patyal}.  The CuBr$_{4}^{-2}$ tetrahedra are co-crystallized 
along with the organic piperidinium 
cations so that the crystal structure resembles a two-leg ladder, Fig. 1.  
The rungs of the ladder are formed 
along the $c^{\ast}$-axis ($19.8^{\circ}$ above the $c$-axis and 
$+23.4^{\circ}$ away from the $a$-$c$ plane \cite{Patyal}) 
by adjacent flattened CuBr$_{4}^{-2}$ tetrahedra 
related by a center of inversion.  
The ladder extends along the $a$-axis with 6.934 {\AA} between Cu$^{2+}$ spins on the same rung 
and 8.597 {\AA} between rungs.
The magnetic exchange, $J_{\perp}$, 
between the Cu$^{2+}$ spins on the same rung is mediated by the orbital overlap of Br 
ions on adjacent Cu sites. The exchange 
along the legs of the ladder, $J_{\parallel}$, is also mediated by somewhat longer 
non-bonding (Br$\cdot \cdot \cdot$Br) contacts and possibly augmented 
by hydrogen bonds to the organic cations.  A frustrating 
diagonal exchange, $J_{F}$, is possible, although it should be weak 
$(J_{F} \ll J_{\parallel})$, and so the potential of a finite $J_{F}$ on the 
short diagonal was considered in our analysis.

\begin{figure}
\centerline{\psfig{figure=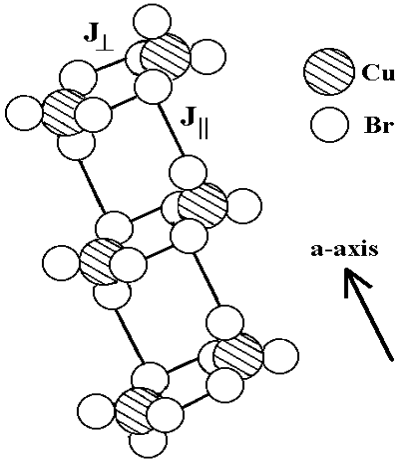,width=3.25in}}
\caption{Schematic of the crystal structure of BPCB.
The legs (rungs) are along the $a$-axis ($c^{\ast}$-axis), see text.}
\label{fig1}
\end{figure}

Shiny, black crystals of BPCB were prepared by slow evaporation of solvent from a methanol solution of 
 [(pipdH)Br] and CuBr$_{2}$, and milling of the 
smallest crystals was used to produce 
the powder samples.  The stochiometry was verified using CHN analysis, and 9 GHz ESR 
measurements were completely consistent with the previously reported data, \emph{i.e.} 
$g$(powder) $= 2.13$ \cite{Patyal}.
In addition, deuterated specimens were produced and used in  
neutron scattering studies 
performed at the HFIR at Oak Ridge National Laboratory.  No evidence for long range 
magnetic order or structural transitions was observed down to 11 K by powder 
diffraction and 1.5 K for single crystal diffraction in the [h 0 l] scattering plane.
The low field magnetic measurements were performed using 
a \textsc{squid} magnetometer.  The high field work was 
conducted at the NHMFL using a 30 T, 33 mm bore resistive magnet and a vibrating (82 Hz)
sample magnetometer equipped with a Cernox thermometer \cite{Brandt}.

The low field, 0.1 T, magnetic susceptibility, $\chi$, of a powder sample, 166.7 mg, 
is shown as a function of temperature in Fig. 2.  The data from single crystals, with 
the magnetic field oriented
along the $a$-, $b$-, and $c$-axes in separate measurements, are indistinguishable from the results 
obtained with the powder specimen.
The general shape of the curve is typical of low dimensional magnetic systems, and more specifically, 
it possesses a rounded peak at $\approx 8$ K and an exponential temperature dependence below the peak.
Consistent with the neutron scattering results, 
no evidence of long-range ordering was observed down to 2 K.  
A small extrinsic Curie-like impurity contribution
(= 1.5\% of the total number of Cu spins) and
a temperature-independent diamagnetic term $(\chi_{dia} = -2.84 \times 
10^{-4}$ emu/mol, which is the sum of 
the core diamagnetism, estimated from Pascal's constants to be 
$-2.64 \times 10^{-4}$ emu/mol, and the background contribution of the sample 
holder)
were subtracted from the data in Fig. 2.  
The Curie-Weiss temperature $\theta$, and the Curie constant, $C$,  
can be extracted from a fit  
$[\chi(T) = \chi_{dia} + C/(T+\theta)$, 50 K $<T<300$ K], and we 
find $C = 0.433 \pm 0.002$ emu K/mol and $\theta = 5.3 \pm 0.1$ K 
\cite{Watson}.   
These values are close to $C=0.425$ emu K/mol ($S=1/2$, $g=2.13$) and 
$\theta \approx (J_{\perp}+2J_{\parallel})/4
= 5.2$ K \cite{Johnston}.      

\begin{figure}
\centerline{\psfig{figure=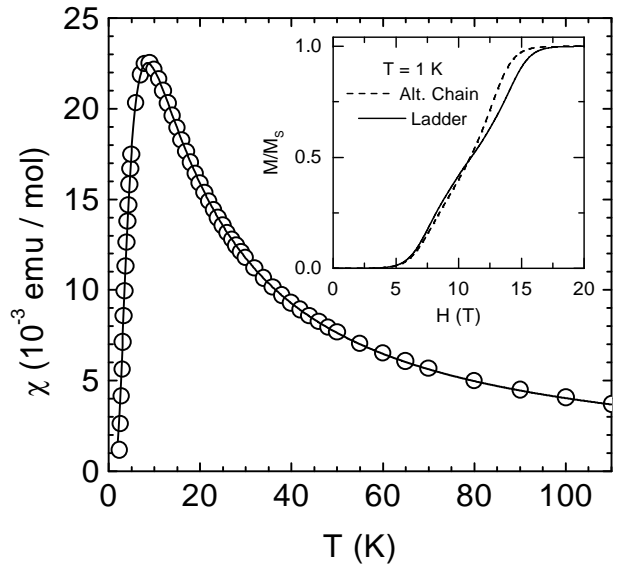,width=3.25in}}
\caption{The $\chi(T)$ of a powder sample (166.7 mg) in 0.1 T.  
The line is 
the result of an exact diagonalization of a ladder Hamiltonian 
with 12 spins when $J_{\perp} = 13.3$ K and $J_{\parallel} = 3.8$ K, see text.  
The inset shows the $M(H,T = 1$ K) expectations of an exact diagonalization of the 
alternating chain and ladder Hamiltonians with the exchange values given in 
the text.}  
\label{fig2}
\end{figure}

Initially, using exact diagonalization methods with 12 spins, 
the $\chi(T)$ data were fit to obtain the values  
$J_{\perp} = 13.3$ K, $J_{\parallel} = 3.8$ K for a ladder Hamiltonian 
and $J_{1} = 13.7$ K , $J_{2} = 5.3$ K  
for an alternating chain Hamiltonian.  
Both fits are 
indistinguishable from the solid line shown in Fig. 2.    
Therefore, using only the low field $\chi(T)$ data, we were unable 
to distinguish between the ladder and alternating chain models, and 
this situation was not 
improved by fitting the $M(H \leq 5$ T, $T = 2$ K) data.  However, 
in extensions up to the saturation magnetization, $M_{s}$, the alternating chain 
model generated $M(H,T<J_{\parallel})$ curves that were asymmetric about 
$M_{s}/2$, as reported for Cu(Hp)Cl \cite{GC,Watson}, and the 
spin ladder description predicted symmetric behavior, see Fig. 2 inset.  
Since our experimental resolution was estimated to be sufficient to 
allow us to differentiate between the two models, 
the high magnetic field studies were initiated.

The high field, $H \leq 30$ T, magnetization of a powder sample, 208.2 mg, is shown in Fig. 3.  Since 
$M_{s}$ was reached in our studies, we were able to directly measure and subtract a small, 
temperature-independent contribution $(\chi_{dia} \approx - 2.84 \times 10^{-4}$ emu/mole), 
which is the same value obtained in our low field work. Measurements were also made on a 
single crystal, 18.9 mg, with $H \parallel a$-axis and for $T \geq 1.6$ K.  Within the 
resolution, the data are the same for the powder and single crystal samples.  
Furthermore, the data were acquired while ramping the field in both directions, and no hysteresis 
was observed.
 
The low energy states of the spin ladder Hamiltonian can be 
mapped, in the strong-coupling limit, 
onto the $S=1/2$ \emph{XXZ} chain \cite{Totsuka}, allowing 
$M(H,T)$ to be modeled. The solid lines in Fig. 3 were 
obtained by numerical integration of the Bethe ansatz equations for the effective 
\emph{XXZ} chain \cite{Takahasi}, using the parameters describing the spin ladder 
fit for $\chi(T)$. All of the data, Figs. 2 and 3, 
are reproduced by one set of exchange values when using the 
ladder model.  On the other hand, the alternating chain model fails to fit 
all of the data with a single set of parameters.  
For example, the 
dashed line in Fig. 3 is $M/M_{s}(H,T=0.7$ K) calculated from 
the alternating chain mapping 
onto the \emph{XXZ} chain when $J_1 = 13.7$ K and $J_2 = 5.3$ K, 
\emph{i.e.} the values obtain from fitting $\chi(T)$, Fig. 2, by an 
alternating chain model.
In addition, our data were analyzed with a ladder model that also 
incorporated a frustrating interaction, $J_{F}$ \cite{Mila}, and 
we can estimate an upper bound of $J_{F} < 0.5$ K.
Consequently, all of the data are consistent with a strongly coupled ladder description for BPCB, 
where $J_{\perp} = 13.3$ $\pm$ 0.2 K, and $J_{\parallel} = 3.8$ $\pm$ 0.1 K.

To leading  order, 
$g \mu_{B} H_{c1} = J_{\perp} - J_{\parallel}$, and 
$g \mu_{B} H_{c2} = J_{\perp} + 2J_{\parallel}$ \cite{Mila,Reigrotzki}.  
Using the previously mentioned parameters, we obtain 
$H_{c1} = 6.6$ T and $H_{c2} = 14.6$ T, \emph{identical} 
with the experimental results.   
The inset in Fig. 3 shows  the derivative curve, 
$d(M/M_{s})/dH$, of our data at the lowest temperature.  
The symmetric double bump structure and its evolution with 
temperature  has been studied
theoretically \cite{Wang} but has not been observed previously in 
$S=1/2$ two-leg ladder materials.  Even though our theoretical curve 
somewhat overestimates the sharpness of $d(M/M_{s})/dH$, the overall 
agreement between theory and experiment, including
the evolution of $M(H,T)$, Fig. 3, is excellent, and 
involves no adjustable parameters once $H_{c1}$ is defined.  
Furthermore, the fact that we see only one feature at $M_{s}/2$ 
between $H_{c1}$ and $H_{c2}$ is evidence that 
our strongly interacting dimers are not coupling to form 
2D \cite{Kageyama} or 3D \cite{Shiramura,Kurniawan} networks. 

\begin{figure}
\centerline{\psfig{figure=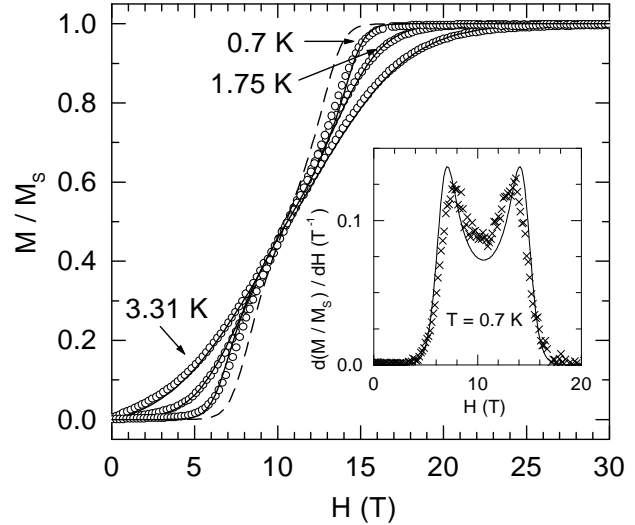,width=3.25in}}
\caption{The normalized magnetization, $M/M_{s}$, of a powder sample (208.2 mg).  
The data traces are limited to $\approx 150$ (of $\approx 3000$) points for clarity.  
The lines are spin ladder predictions of 
an effective \emph{XXZ} chain when $J_{\perp} = 13.3$ K and $J_{\parallel} = 3.8$ K.
At $T = 0.7$ K, the inflection point at $M_{s}/2$ is clearly visible, and the 
inset shows 
the derivative of this data.  The dashed line is the alternating chain model 
prediction for 0.7 K when  
$J_{1}$, $J_{2}$ are taken to be the values obtained from fitting the data in Fig. 2, see text.}
\label{fig3}
\end{figure} 

At $H_{c1}$, BPCB undergoes a transition from gapped dimer pairs 
to  a gapless Luttinger liquid phase with fermionic excitations, 
where the magnetization is proportional to the fermion density 
\cite{Sachdev2,Tsvelik,Affleck}. This transition can be described as  a 
condensation of a dilute gas of bosons (dimers), and quasiparticle interactions are
irrelevant at the transition point. At $H_{c2}$, an analogous situation exists where 
the transition is between the Luttinger liquid and spin polarized phases.  
When $T$, $g\mu_{B}|H-H_{c1}|$, and $g\mu_{B}|H_{c2}-H|$ are $\lesssim 
J_{\parallel}$, the 1D magnetization is predicted to obey 
the universal scaling law (assuming $J_{\perp}/J_{\parallel} \gg 1$) that may be written as: 
\[
\frac{M(H,T)}{M_{s}}=\sqrt{2k_{B}T/J_{\parallel}} \: {\cal M}(g \mu_{B} [H - H_{c1}]/k_{B}T)\;,
\]
\[
1-\frac{M(H,T)}{M_{s}}=\sqrt{2k_{B}T/J_{\parallel}} \: {\cal M}(g \mu_{B} [H_{c2} - H]/k_{B}T)\;, 
\]
where the universal function ${\cal M}$ is the fermion density \cite{Sachdev2}.  
This theoretically predicted scaling behavior is 
compared to the data in Fig. 4, where the agreement is 
impressive.  It is important to stress that the scaling shown in Fig. 4 has been theoretically 
predicted \cite{Sachdev2} and is not a result of extracting scaling variables on the basis of the 
data \cite{Chaboussant4}.  In an isolated spin ladder, scaling is expected at the lowest 
temperatures, $T \lesssim J_{\parallel}$. A deviation from scaling is observed for $T=0.7$ K, which 
suggests that other weak interactions, such as $J_{F}$ or $J^{\prime}$, may begin to have a subtle 
influence, while the data up to 4.47 K appear to obey 
the scaling theory.  The $T^{1/2}$ scaling of the magnetization 
at the critical point $H=H_{c1}$, Fig. 4, is 
further evidence that BPCB is a two-leg spin ladder with $J^{\prime} \ll J_{\parallel}$ 
\cite{Giamarchi}.        

\begin{figure}
\centerline{\psfig{figure=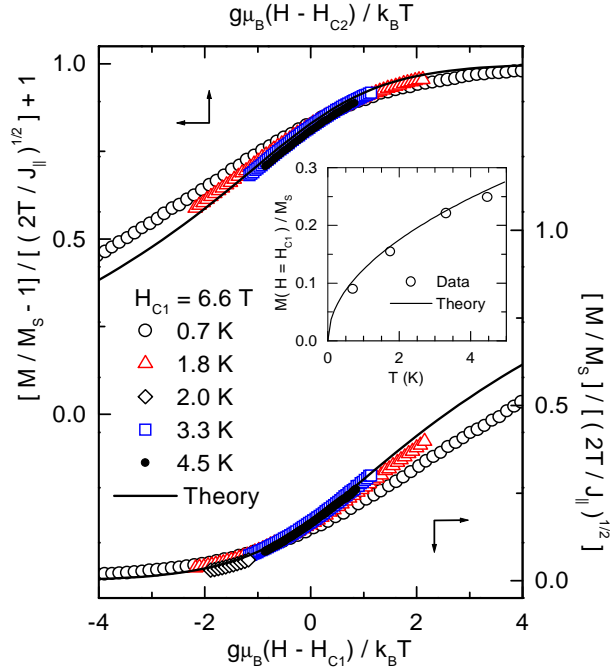,width=3.25in}}
\caption{The scaled data in the vicinity of $H_{c1}$ and $H_{c2}$.  
The solid lines are the 
predictions of the theory when $H_{c1}$ is fixed.  The inset shows the $T^{1/2}$ scaling 
behavior at $H_{c1}$.} 
\label{fig4}
\end{figure}

In summary, analysis of $M(H \leq 30$ T, $T \geq 0.7$ K) has allowed us to identify BPCB 
as a two-leg $S=1/2$ spin ladder in the strong-coupling limit, $J_{\perp}/J_{\parallel} \sim  3.5$.  
A single set of exchange constants, $J_{\perp} = 13.3$ K and $J_{\parallel} = 3.8$ K, 
are able to accurately describe all of the data. 
The $M(H \approx H_{c1}$ or $H_{c2}, 1$ K $< T < 4.5$ K) data exhibit  scaling behavior
in the  universality class of the 1D dilute Bose gas transition \cite{Sachdev1,Sachdev2}.  
Although we have  considered the potential existence of additional  exchange 
interactions $J_{F}$ and $J^{\prime}$, effects arising from these parameters are not prominent 
in the present data.  However, since subtle differences arise between the theoretical 
predictions and the data at the lowest temperature, additional perturbing interactions  
may be present.  

We have enjoyed input from many colleagues, 
including J. H. Barry, A. Feher,  M. Orend\'{a}\v{c}, 
A. Orend\'{a}\v{c}ova, F. Mila, and A. Yashenkin.
We thank G. Chaboussant for sending Ref. \cite{GC}.  
This work was supported, in part, by the NSF through DMR-9704225 (B.C.W. and M.W.M.), 
DMR-9357474 (V.N.K.), the NHMFL via DMR-9527035, DMR-9900855 (D.R.T. and coworkers), 
and by the State  of Florida.  
Oak Ridge National Laboratory is managed for the DOE by 
UT-Battelle, LLC, under Contract No. DE-AC05-00OR22725.

\end{document}